%% file: main.tex
\documentclass[sigconf]{acmart}

\AtBeginDocument{%
  \providecommand\BibTeX{{%
    \normalfont B\kern-0.5em{\scshape i\kern-0.25em b}\kern-0.8em\TeX}}}

\usepackage{tikz}
\usepackage{subfigure}
\usepackage{float}
\usepackage{multirow}
\usepackage{caption}

\usepackage{multirow}
\newcommand{\ouralg}{HDLock}

\setcopyright{none}
\renewcommand\footnotetextcopyrightpermission[1]{}
\begin{document}

\title{\ouralg: Exploiting Privileged Encoding to Protect Hyperdimensional Computing Models against IP Stealing} 

\author{Shijin Duan}
\email{duan.s@northeastern.edu}
\affiliation{
 \institution{Northeastern University}
 \city{Boston}
 \state{MA}
 \country{USA}
}

\author{Shaolei Ren}
\email{sren@ece.ucr.edu}
\affiliation{
 \institution{UC Riverside}
 \city{Riverside}
 \state{CA}
 \country{USA}
}

\author{Xiaolin Xu}
\email{x.xu@northeastern.edu}
\affiliation{
 \institution{Northeastern University}
 \city{Boston}
 \state{MA}
 \country{USA}
}

\begin{abstract}
Hyperdimensional Computing (HDC) is facing infringement issues due to straightforward computations. This work, for the first time, raises a critical vulnerability of HDC --- an attacker can reverse engineer the entire model, only requiring the unindexed hypervector memory. To mitigate this attack, we propose a defense strategy, namely \ouralg, which significantly increases the reasoning cost of encoding. Specifically, \ouralg\ adds extra feature hypervector combination and permutation in the encoding module. Compared to the standard HDC model, a two-layer-key \ouralg\ can increase the adversarial reasoning complexity by 10 order of magnitudes without inference accuracy loss, with only 21\% latency overhead.
\end{abstract}

\maketitle

\section{Introduction} \label{sec:Introduction}
As an alternative to the deep neural networks (DNNs), brain-inspired hyperdimensional computing (HDC) is proposed as a promising solution to classification tasks with higher efficiency and less storage footprint \cite{kanerva2009hyperdimensional}. For example, the recent quantized HDC \cite{imani2019quanthd} inference can be 10x faster than that of a binary neural network at the same model size and accuracy. Although not designed for complex learning tasks, HDC is particularly suitable for real-time classification on resource-limited devices, such as lightweight Internet of Things (IoT) devices and wearables.

The ultra efficiency and lightweight nature of HDC has attracted many research interests. As a result, HDC implementations have been explored for different hardware platforms, such as FPGA \cite{imani2019quanthd}, GPU \cite{kim2020geniehd}, and in-memory-computing systems \cite{karunaratne2020memory}. Unfortunately, compared to the emerging studies on the performance improvement of HDC algorithm, its security is significantly under-explored. Few existing works have explored the attack \cite{yang2020adversarial} and defense \cite{khaleghi2020prive} of the HDC inference with a focus on input, leaving the HDC model security under-explored. Similar as other machine learning methods, for which the model intellectual property (IP) is of high confidentiality \cite{liu2018survey}, the HDC models should also be well preserved against IP stealing or model extraction attacks. Building a high-performance HDC model involves multiple stages, including expensive training data collection as well as careful hyperparameter tuning for class hypervector construction (e.g.,  the number of retraining rounds and ``learning rate'' \cite{imani2019quanthd}). As a result, it is very costly to produce a well-performing HDC model, which could become an target of IP stealing or model extraction attacks. 

This work, for the first time, explores the IP security of HDC models and raises one critical vulnerability on the current HDC encoding module, which could leak the entire HDC model. This vulnerability is associated with the unique summation and multiplication structure in the encoding phase of HDC models. Leveraging such vulnerability, the attacker could extract the entire encoding module, even without the knowledge of the mapping information of hypervectors. As a result, the attacker can steal the HDC model or even reason the training dataset. Further, we propose a novel defense framework, namely \ouralg, to mitigate such vulnerability. \ouralg\ constructively applies combination and permutation to the HDC encoding module to protect its model IP. Specifically, the feature hypervectors are derived from several selected and permuted base hypervectors, making it significantly challenging for the attacker to reason the feature hypervectors. The base hypervector selection and permutation rules are regulated by a key, which can be stored in a tamper-proof memory as many circuit locking schemes \cite{xie2016mitigating}. \ouralg\ can efficiently mitigate the HDC IP stealing attacks, disabling the attacker from acquiring the encoding details without correct guess on the feature hypervectors. 

The main contributions of this work are as follows:
\begin{itemize}
    \item To the best of our knowledge, this is the first work investigating the IP security of the HDC model. Leveraging the raised vulnerability, attackers can craft specific adversarial inputs and quickly reason the mapping information by observing the encoding outputs.
    
    \item As mitigation, we propose a defense framework, \ouralg, to significantly increase the adversarial reasoning cost (complexity). \ouralg\ constructively uses combination and permutation to protect the HDC model, which effectively eliminates reasoning attack while still being hardware friendly. 
    
    \item We thoroughly evaluate the vulnerability and the proposed framework. Experimental results demonstrate that \ouralg\ can significantly harden the adversarial reasoning. For example, the reasoning complexity can be enlarged by 10 order of magnitudes compared to the original model with only 21\% time overhead, without the inference accuracy loss.
\end{itemize}

The remainder of this paper is organized as follows. Sec. \ref{sec:Background} briefly reviews the preliminaries of HDC. Sec. \ref{sec:Vulnerability} presents the the discovered vulnerability of HDC models in detail. Sec. \ref{sec:Strategy} illustrates the proposed defense framework \ouralg, the evaluations and results on the vulnerability and countermeasure are presented in Sec. \ref{sec:Evaluation}. Sec. \ref{sec:Conclusion} concludes this paper.

\section{Preliminaries of HDC} \label{sec:Background}

Hyperdimensional Computing (HDC) is a paradigm that represents object feature indices and values using hyperdimensional vectors (HV), $HV\in \{1,-1\}^D$ \cite{kanerva2009hyperdimensional}. Specially, the hypervectors representing feature indices ($FeaHV$) are supposed to be orthogonal to each other, while the hypervectors representing feature values ($ValHV$) are usually linearly correlated, corresponding to the feature value correlation \cite{yang2020adversarial}. The normalized Hamming distance is used to differentiate object $FeaHV$s and $ValHV$s, as follows:
\begin{subequations}
\begin{align}
    Hamm(FeaHV_{i_1},FeaHV_{i_2}) &\approx 0.5 \label{eq:Hamm_fea}\\
    Hamm(ValHV_{v_1},ValHV_{v_2}) &\approx 0.5\times \frac{|v_1-v_2|}{v_{max}-v_{min}} \label{eq:Hamm_val}
\end{align}
\end{subequations}
where $i_1$ and $i_2$ are two arbitrary feature indices, and $v_1, v_2$ are two example values in the value range $[v_{min}, v_{max}]$.

The most commonly used operators in HDC are Multiplication-Addition-Permutation (MAP). Taking two hypervectors $HV_1$ and $HV_2$ in the bipolar system (i.e., $\{1,-1\}^D$) as an example, these three operations can be represented as $HV_1\times HV_2$, $HV_1+HV_2$, and $\rho(HV_1)$, respectively. Multiplication and addition are computed in element-wise, while permutation generates a new hypervector dissimilar from the original one in a customized manner. The simplest permutation is to circularly rotate the hypervector by certain elements, which is commonly used in current HDC works. For example, $\rho^k(HV_1) = \{HV_1[k:D-1],HV_1[0:k-1]\}$ means rotating $HV_1$ by $k$ bits.

\begin{figure}[!t]
  \centering
  \includegraphics[width=0.9\linewidth]{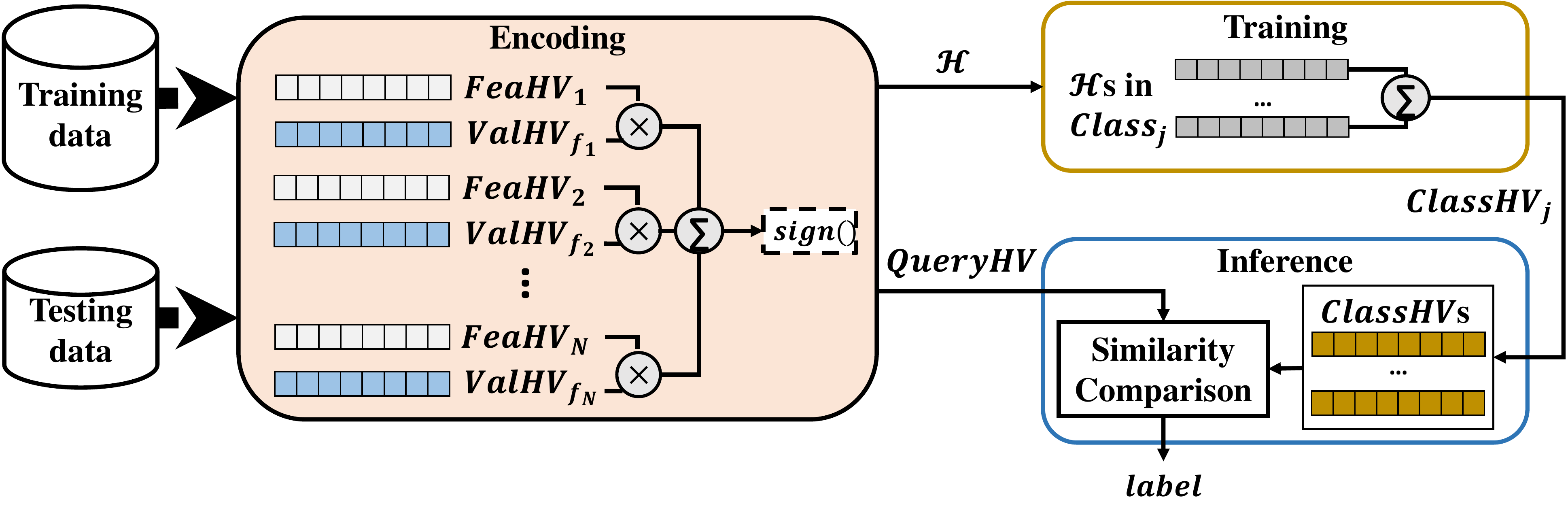}
  \caption{The overview of HDC classification model.}
  \label{fig:basic_HDC}
\end{figure}

One emerging application of HDC is the classification tasks. As illustrated in Fig. \ref{fig:basic_HDC}, the HDC-based classification consists of three steps: \textit{Encoding}, \textit{Training}, and \textit{Inference}.

\paragraph{Encoding:}
An input sample of the HDC encoding module is denoted as a feature vector ($\textbf{F}= \{f_1, f_2, ..., f_N\}$) with $N$ features. The feature values are discretized to $M$ levels based on the minimum and maximum values across the entire dataset, for which $M$ correlated value hypervectors are constructed \cite{imani2019quanthd}. The encoding module will represent the input feature $\textbf{F}$ with a $D$-dimensional hypervector. Specifically, $N$ orthogonal $FeaHV$s and $M$ consecutive $ValHV$s are generated, corresponding to $N$ feature indices and $M$ possible discretized values. For non-binary HDC, Encoding adds up all the feature hypervectors and value hypervectors as output $\mathcal{H}_{nb}$:
\begin{equation}
    \mathcal{H}_{nb} = \sum^N_{i=1} ValHV_{f_i}\times FeaHV_i
    \label{eq:non_binary_HDC}
\end{equation}
 
To achiever higher computing-efficiency, the $\mathcal{H}_{nb}$ can be binarized as $\mathcal{H}_{b}$, in binary HDC:
\begin{equation}
\small
    \mathcal{H}_{b} = sign\left(\sum^N_{i=1} ValHV_{f_i}\times FeaHV_i\right) = sign(\mathcal{H}_{nb})
    \label{eq:binary_HDC}
\end{equation}
where $sign(\cdot)$ denotes the sign (binarizing) function while $sign(0)$ is randomly assigned to $-1$ or $1$. 

\paragraph{Training:}
With the input sample represented by hypervector $\mathcal{H}$, a non-binary HDC model can be trained as 
\begin{equation}
\small
    ClassHV_j = \sum_{\mathcal{H} \in \Omega_j} \mathcal{H}
\end{equation}
where $\Omega_j$ represents the hypervectors of the $j$-th object class assuming there are $C$ classes in total. Similarly, the result can also be binarized in the binary HDC model.

\paragraph{Inference:}
For inference, a query sample is firstly encoded to a hypervector $QueryHV$, using the same feature and value hypervectors. Then, the similarity between $QueryHV$ and the $ClassHV$s is calculated, and the most similar $ClassHV$ indicates the inference label. The similarity between the $QueryHV$ and $ClassHV$ is quantified using \textit{cosine} function for non-binary model, which calculates the angle between these two hyeprvectors. In binary model, Hamming distance is used as all hypervectors are binary \cite{kleyko2018classification}.

\section{HDC Model IP Vulnerability} \label{sec:Vulnerability}
Considering the intellectual property (IP) value of the HDC model, its base hypervectors (e.g. feature and value hypervectors in Fig. \ref{fig:basic_HDC}) should be well-protected. Otherwise, once the attacker learns the base hypervectors, s/he can easily duplicate a similar HDC model for malicious attacks, such as reverse engineering the inputs \cite{khaleghi2020prive} or generating adversarial inputs \cite{yang2020adversarial}. Moreover, due to the simple composition in an HDC model, it is very easy to be reverse engineered, as detailed below. 

\subsection{Threat Model} 
To enable efficient inference, HDC models are typically run on resource-constrained in-memory computing platforms or FPGAs.
Unfortunately, most existing security solutions could not properly used for protect the HDC models on these platforms. For example, the physical isolation techniques \cite{corbett2013xilinx} proposed for FPGA security have been tampered with practical attacks \cite{zhao2018fpga}, while the security consideration on in-memory computing is under-explored. 

As the first work exploring HDC model IP security, we use a strong threat model, in which the HDC model owner protects the model IP by keeping users from directly accessing the encoding module details. For example, the IP owner stores the index mapping\footnote{Index mapping represents the mapping information between features/values and corresponding hypervectors.} of the base hypervectors in a secure environment, such as a tamper-proof memory preventing from probing internal signals, as suggested in \cite{xie2016mitigating}. Considering the limited computing resources in lightweight IoT devices, we assume that there are insufficient secure memory to store the entire HDC model (usually of MegaBytes) for IP protection purpose. Therefore, protecting the index mapping is an adequate solution for IP protection, which is much more memory-efficient than protecting all the hypervectors in the presence of tiny secure memory space.

Consequently, our threat model assumes that the raw data of base hypervectors are vulnerable while the mapping information is preserved from attacker access. As a result, the attacker can only access the unindexed hypervectors stored in the non-secured memory, and craft his/her own inputs and observe the encoding outputs. Note that our threat model is more strict than the white-box assumption in \cite{khaleghi2020prive}, since the hypervectors are stored publicly while the mapping information (as a ``key'') can be stored in secure memory to lock the encoding module. 

\begin{figure*}
\centering
\begin{minipage}{.65\textwidth}
  \centering
  \includegraphics[width=\linewidth]{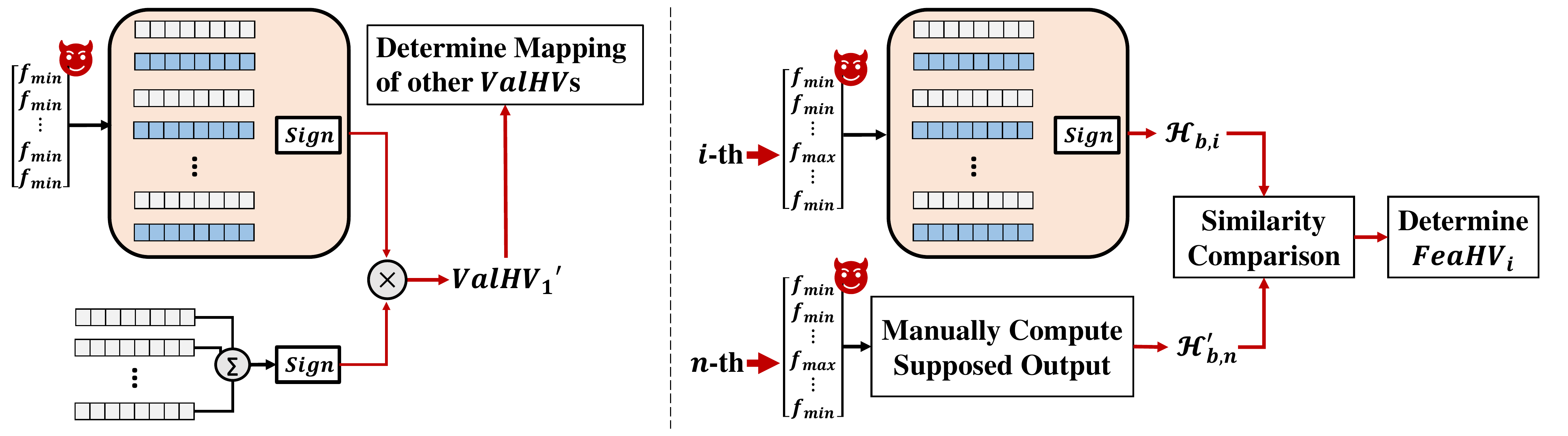}
  \caption{Attack flow for the encoding of binary HDC.}
  \label{fig:attack_record}
\end{minipage}%
\begin{minipage}{.3\textwidth}
  \centering
  \includegraphics[width=0.8\linewidth]{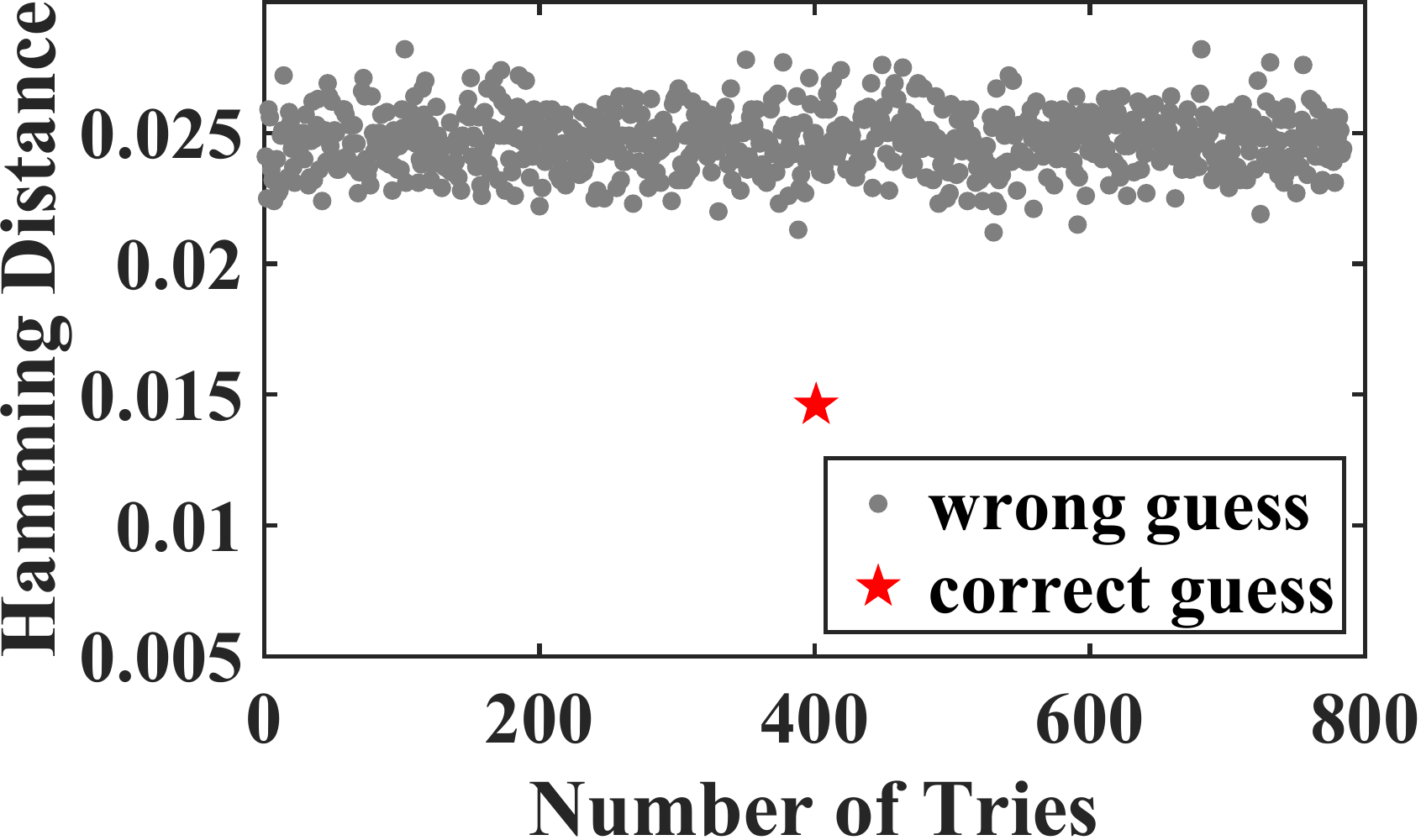}
  \caption{The Hamming distances between the guesses and the ground truth on the 784 possible guesses.}
  \label{fig:attack_normal_pixel}
\end{minipage}
\end{figure*}

\subsection{Reasoning HDC Model with Divide-And-Conquer Strategy}
This section presents the vulnerability of the HDC models, which can be utilized by the adversary to deduce the mapping information, even only given the base hypervectors (i.e., the unindexed $FeaHV$s and $ValHV$s). We use the same notations presented in Sec. \ref{sec:Background}, that the encoding input has $N$ features and $M$ discretized value levels, corresponding to $N$ orthogonal $FeaHV$s and $M$ consecutively distributed $ValHV$s, where $ValHV_1\perp ValHV_M$. 

Since the encoding entangles the $FeaHV$s and $ValHV$s with multiplication, the hypervector mapping information cannot be directly derived from the encoding output. Although protecting the mapping information can resist brute force attack, we demonstrate that it is still vulnerable to our proposed attack flow, as shown in Fig. \ref{fig:attack_record}. Since the non-binary HDC model is vulnerable as it has no obfuscation operation (i.e., the binarization in binary HDC) to compress the information in hypervectors, thus we use the more-secure binary HDC model as a case-study.
The proposed attack strategy can be conducted in the following two steps:

\textbf{Value Hypervector Extraction:}
The inherent weakness of $ValHV$s lies in the consecutive distribution, which means only $ValHV_1$ and $ValHV_M$ are orthogonal while all the other $ValHV$s are uniformly distributed in between, as shown in Eq. \ref{eq:Hamm_val}. Hence, the two farthest hypervectors, $ValHV_1$ and $ValHV_M$, can be determined by observing the Hamming distances between all value hypervectors. To identify these two $ValHV$s, attacker can craft an adversarial sample, of which all the features have the minimum value (i.e., corresponding to $ValHV_1$), so the encoding output is
\vspace{-5pt}
\begin{equation}
\begin{split}
\footnotesize
    \mathcal{H}_{b,min\_value} &= sign\left(\sum_{i=1}^{N} FeaHV_i \times ValHV_1\right)\\
    &= ValHV_1\times sign\left(\sum_{i=1}^{N} FeaHV_i\right)
\end{split}
\end{equation}
One property for the single-value input is that the $ValHV$ can be moved out, so the summation of $FeaHV$s is treated as a whole without consideration of the inner arrangement. 
Therefore, the attacker can simply estimate the $ValHV_1$ 
\begin{equation}
\small
    ValHV_1 ' = \mathcal{H}_{b,min\_value}\times sign\left(\sum_{i=1}^{N} FeaHV_i\right)
\label{eq:reason_ValHV}
\end{equation}
By comparing the similarity between the estimated $ValHV_1'$ and the two $ValHV$ candidates, the attacker can determine the correct $ValHV_1$. The mapping information of $ValHV_M$ and other $ValHV$s can also be determined subsequently.

\textbf{Feature Hypervector Extraction:}
With the mapping of value hypervectors, the adversary can deduce the $FeaHV$ mapping using specific inputs. Taking the first feature hypervector $FeaHV_1$ as an example, the attacker can craft an adversarial input, in which the first feature value is the maximum ($ValHV_M$) and other feature values are the minimum ($ValHV_1$). Hence, the output $\mathcal{H}_{b,1}$ can be denoted as: 
\begin{equation}
\footnotesize
    \mathcal{H}_{b,1} = sign\left(FeaHV_1 \times ValHV_M + \sum^{N}_{i=2} FeaHV_i \times ValHV_1\right)
\end{equation}
Using such crafted inputs, the attacker can separate and analyze the feature of interest from the summation individually. Since all the value hypervectors are determined, the mapping information of $FeaHV$s can be deduced with the divide-and-conquer strategy, i.e., the attacker assumes the $n$-th feature hypervector in the candidate pool corresponds to the real $FeaHV_1$, and denotes it as $FeaHV_n$. S/he constructs another encoding module and calculates the output with the selected $FeaHV_n$:
\begin{equation}
\footnotesize
    \mathcal{H}_{b,1}' = sign\left(FeaHV_n \times ValHV_M + \sum^{N}_{i=1, i\neq n} FeaHV_i \times ValHV_1\right)
\end{equation}
By iterating all candidates in the feature hypervectors pool, the attacker finds one feature hypervector with which the derived $\mathcal{H}_{b,1}'$ is closest (i.e., smallest Hamming distance) to the $\mathcal{H}_{b,1}$. Following this strategy, the attacker can determine the mapping information for the feature hypervectors. The computing complexity is $O(N^2)$, since the divide-and-conquer method divides the permutation into $N$ independent tasks and solves them one by one. 

To explicitly demonstrate the feasibility of the proposed attack strategy, we use the HDC model of MNIST as an example. To reason the feature hypervectors, we craft an adversarial input image to attack the first pixel, in which the first pixel is set as white (255) and all other pixels are set as black (0). As proof-of-concept, we set the 400-th feature hypervector ($FeaHV_{400}$) as the one corresponding to the first pixel, i.e., the correct guess. The result in Fig. \ref{fig:attack_normal_pixel} shows that the correct guess generates an $\mathcal{H}_{b,1}'$ with much lower Hamming distance to the golden $\mathcal{H}_{b,1}$ than other wrong guesses. 

This attack flow can be extended to the non-binary HDC encoding module, for which the difference between correct and wrong guesses is larger, thus the correct guess will make the $cosine$ value exactly be 1 with a 100\% confidence. This vulnerability can be easily explored to make the HDC model IP infringed. Since encoding module is the critical component for all HDC models, this vulnerability calls for more attention to protect the HDC encoding module. 

\section{HDC Model Locking as a Defense} \label{sec:Strategy}
In this section, we present a resource-friendly model locking framework for the encoding modules, namely \ouralg, for both binary and non-binary HDC models. \ouralg\ significantly increases the searching cost of the raised attacks, making it infeasible for the attacker to reason the mapping information within an acceptable time duration. Meanwhile, we avoid complicated locking designs that suffer in efficiency and performance. 

\subsection{\ouralg\ Framework Overview}
\begin{figure}[!t]
  \centering
  \includegraphics[width=0.8\linewidth]{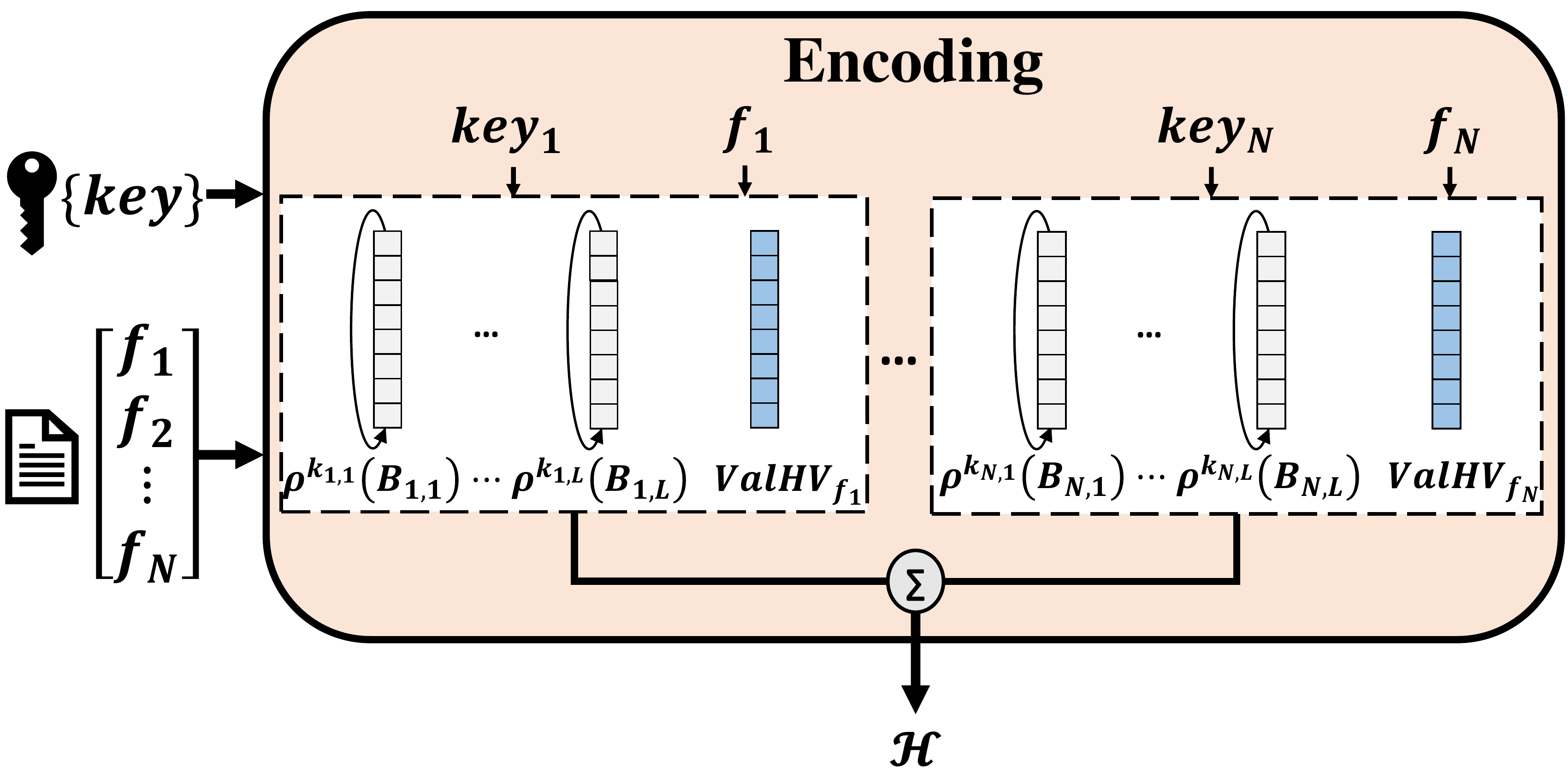}
  \caption{The encoding module of the \ouralg\ framework.}
  \label{fig:gen_stra}
\end{figure}

To significantly enlarge the search space for the attacker, \ouralg\ modifies the calculation of encoding module, and still keeps the mapping information as a key to lock the HDC model. Fig. \ref{fig:gen_stra} presents the \ouralg\ framework on the encoding module.

In \ouralg, the feature hypervector is represented as the product of $L$ permuted base hypervectors:
\begin{equation}
\small
    FeaHV_i = \prod^L_{l = 1} \rho^{k_{i,l}}(B_{i,l})
    \label{eq:encryption}
\end{equation}
where $L$ stands for the layers of representation, and $\rho^{k_{i,l}}(\cdot)$ is the permutation that rotates the $l$-th base hypervector on the $i$-th feature by $k_{i,l}$ bits. Here, the base hypervectors ($B$s) are randomly generated and orthogonal to each other. We assume the generated pool has $P$ base hypervectors stored in the public memor, and the indices and permutation value is stored in the secure memory as the key of the $FeaHV$s generation in Eq. \ref{eq:encryption}. Regarding to the generated $FeaHV$s, the encoding output with \ouralg\ framework $\mathcal{H}_{Lock}$ is
\begin{equation}
\small
\begin{split}
    \mathcal{H}_{Lock} &= \sum^N_{i=1} FeaHV_i\times ValHV_{f_i} \\
    &= \sum^N_{i=1} \left (ValHV_{f_i}\times \prod^L_{l = 1} \rho^{k_{i,l}}(B_{i,l})\right )
\end{split}
\label{eq:comb_HDC}    
\end{equation}
The key will store $N\times L$ base hypervector mapping information. Specifically, the key is composed of $N$ sub-keys, where $key_i$ is applied to the $i$-th feature constructed by $L$ permuted base hypervectors. $k_{i,l}$ and $index(B_{i,l})$ show the rotated bits and the base hypervector index for the $l$-th hypervector on the $i$-th feature. Given a wrong guess of the key, the encoding output $\mathcal{H}_{Lock}$ will also be wrong. Even using the proposed divide-and-conquer attacking strategy, the complexity of figuring out all the $FeaHV$s is $O\left(N\cdot(DP)^{L}\right)$.

\paragraph{Why Not Represent the Value Hypervectors?}
In the proposed \ouralg\ framework, only the feature hypervectors are represented with combination and permutation on base hypervectors, while the value hypervectors are still open to access. If $ValHV$s are constructed by base hypervector combination, the base hypervectors must be correlated due to the correlation of $ValHV$s, which greatly weakens the resistance against the reasoning attacks. Further, the design of base hypervectors for $ValHV$s is complicated, in order to retain the linear correlation. Therefore, jointly considering the security, effectiveness, and resource/time overhead, we find that only protecting $FeaHV$s is sufficient if the required reasoning time is already unaffordable to the attackers.

\subsection{Security Validation}
In this section, we validate the security of $FeaHV$s represented with combination and permutation. Since $ValHV$s are not protected, we assume a strong attack model in which the attacker already obtained the entire $ValHV$s mapping information, i.e., only the mapping of $FeaHV$s needs to be reasoned.

By attacking one $FeaHV$, more tricky strategies are required to reason the correct mapping. Taking the first feature $FeaHV_1$ as an example, where two adversarial inputs can be generated: the first one has all features of the minimum value ($ValHV_1$), while the other is the same except that its first feature has the maximum value ($ValHV_M$). Hence, these two encoding outputs $\mathcal{H}_{Lock}^1$ and $\mathcal{H}_{Lock}^M$ are
\begin{equation}
\footnotesize
\begin{split}
&\mathcal{H}_{Lock}^1 = sign\left( ValHV_1\times \prod^L_{l = 1} \rho^{k_{1,l}}(B_{1,l}) + \mathcal{H}_0 \right)\\
&\mathcal{H}_{Lock}^M = sign\left( ValHV_M\times \prod^L_{l = 1} \rho^{k_{1,l}}(B_{1,l}) + \mathcal{H}_0 \right)
\end{split}
\label{eq:lock_attack}
\end{equation}
where
\begin{equation}
\footnotesize
    \mathcal{H}_0 = \sum^N_{i=2} \left (ValHV_1\times \prod^L_{l = 1} \rho^{k_{i,l}}(B_{i,l})\right )
\end{equation}
is the constant part. Therefore, the difference between $\mathcal{H}_{Lock}^1$ and $\mathcal{H}_{Lock}^M$ is totally resulted in by the difference on the first term. Although $\mathcal{H}_0$ dominates the output, there are still a few different elements caused by the first term. Leveraging this information, the attacker can determine which guess is correct.

To create a criterion judging the guesses, the attacker does subtraction on the two outputs in Eq. \ref{eq:lock_attack}, and select the indices $\mathcal{I}$ whose elements are non-zero. Afterwards, s/he generates a guess $\prod^L_{l = 1} \rho^{k_{g,l}}(B_{g,l})$, and calculate
\begin{equation}
\footnotesize
    \mathcal{H}_{attack}=sign\left(\left(ValHV_1 - ValHV_M\right)\times \prod^L_{l = 1} \rho^{k_{g,l}}(B_{g,l})\right)
\label{eq:attack_vector}
\end{equation}
where $k_{g,l}$ and $index(B_{g,l})$ denote the guesses on the permutation and index of base hypervector, respectively. The attacker can calculate Hamming distance of the subtraction result and $\mathcal{H}_{attack}$ on indices $\mathcal{I}$. The correct guess of the key on the first feature has the lowest Hamming distance. To reason one feature mapping, the attack needs $(DP)^{L}$ guesses.

\begin{figure*}
\begin{minipage}{0.3\textwidth}
\centering
\subfigure[{\scriptsize Attack on $k_{1,1}$}]{
        \centering
        \label{fig:perm1_bin}
		\includegraphics[width=0.48\linewidth]{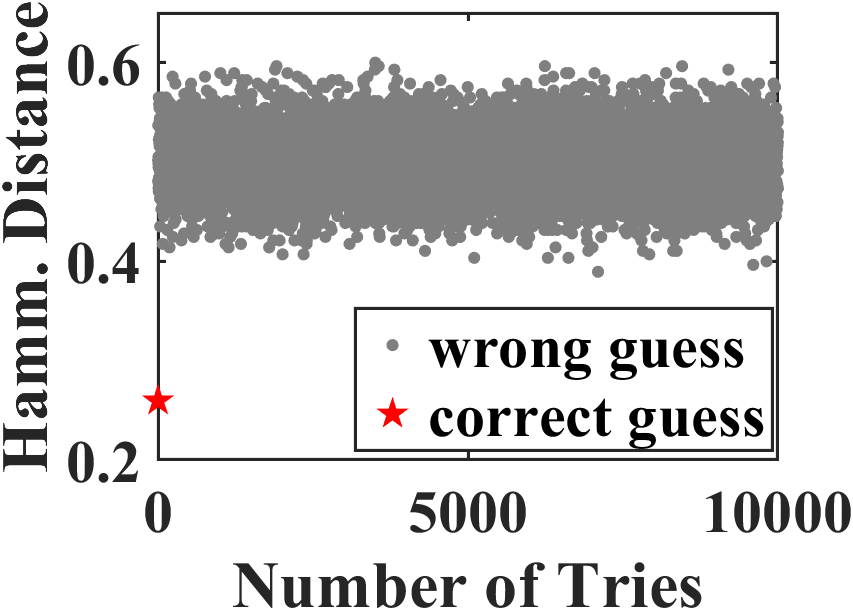}}
\subfigure[{\scriptsize Attack on $index(B_{1,1})$}]{
        \centering
        \label{fig:comb1_bin}
		\includegraphics[width=0.44\linewidth]{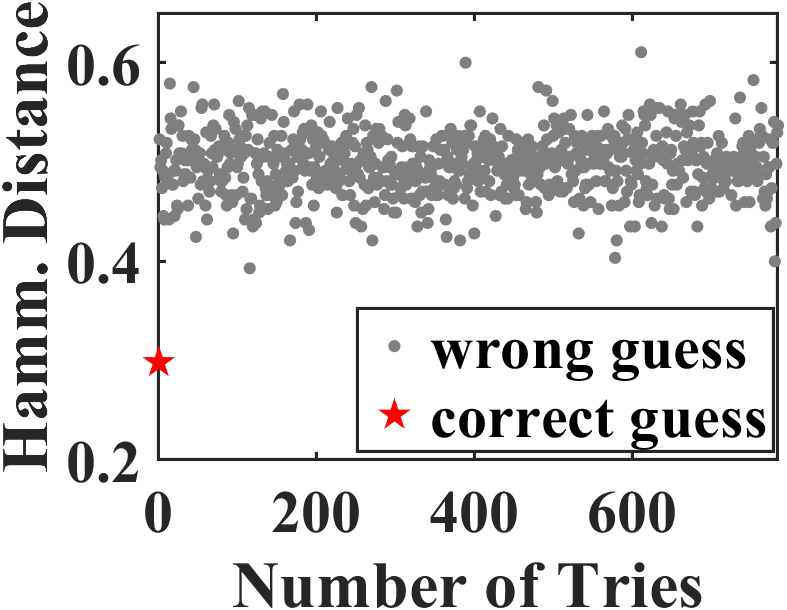}}
\subfigure[{\scriptsize Attack on $k_{1,2}$}]{
        \centering
        \label{fig:perm2_bin}
		\includegraphics[width=0.48\linewidth]{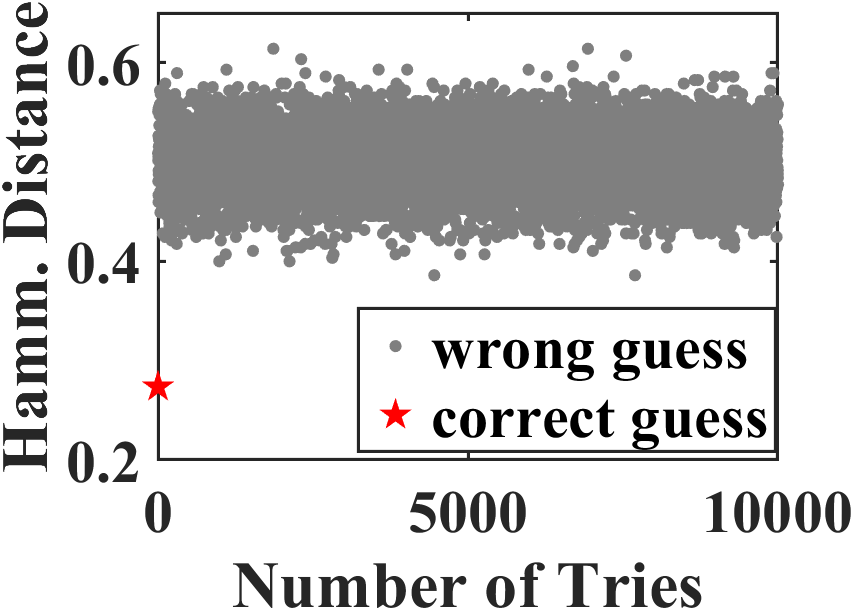}}
\subfigure[{\scriptsize Attack on $index(B_{1,2})$}]{
        \centering
        \label{fig:comb2_bin}
		\includegraphics[width=0.44\linewidth]{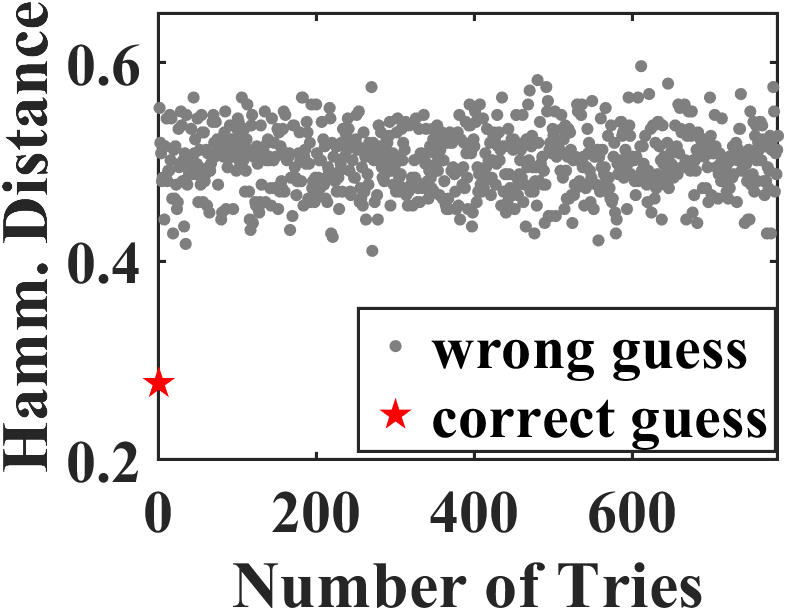}}
\caption{ \ouralg\ security validation on binary HDC.}
\label{fig:validation_binary}
\end{minipage}%
\hspace{0.02\textwidth}
\begin{minipage}{0.3\textwidth}
\centering
\subfigure[{\scriptsize Attack on $k_{1,1}$}]{
        \centering
        \label{fig:perm1_nonbin}
		\includegraphics[width=0.48\linewidth]{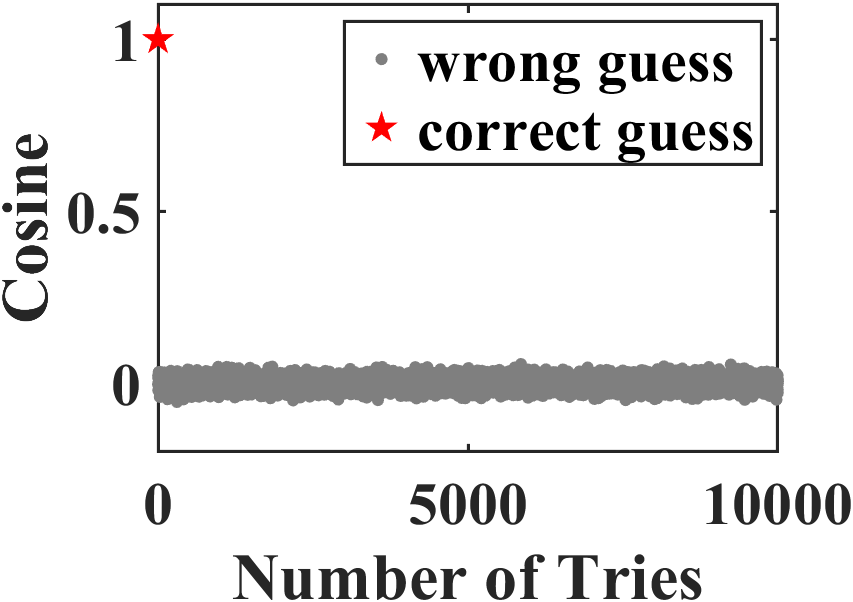}}
\subfigure[{\scriptsize Attack on $index(B_{1,1})$}]{
        \centering
        \label{fig:comb1_nonbin}
		\includegraphics[width=0.44\linewidth]{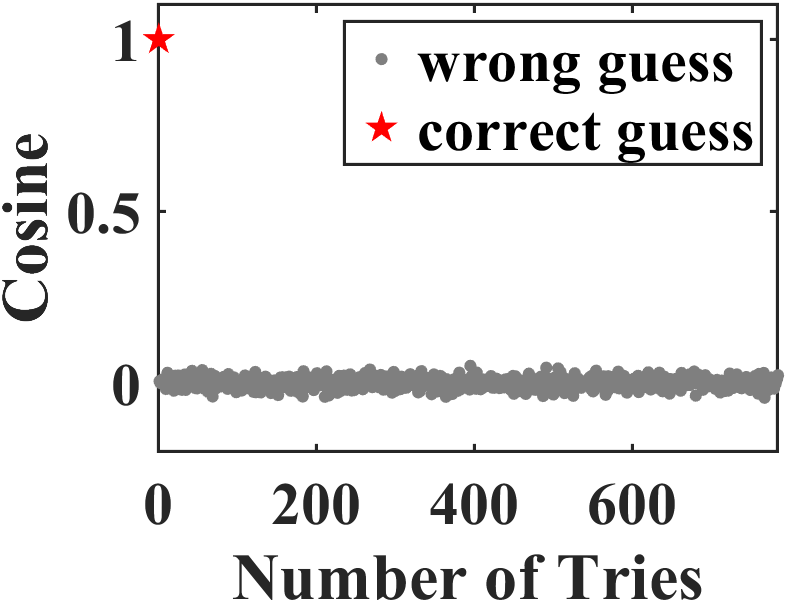}}
\subfigure[{\scriptsize Attack on $k_{1,2}$}]{
        \centering
        \label{fig:perm2_nonbin}
		\includegraphics[width=0.48\linewidth]{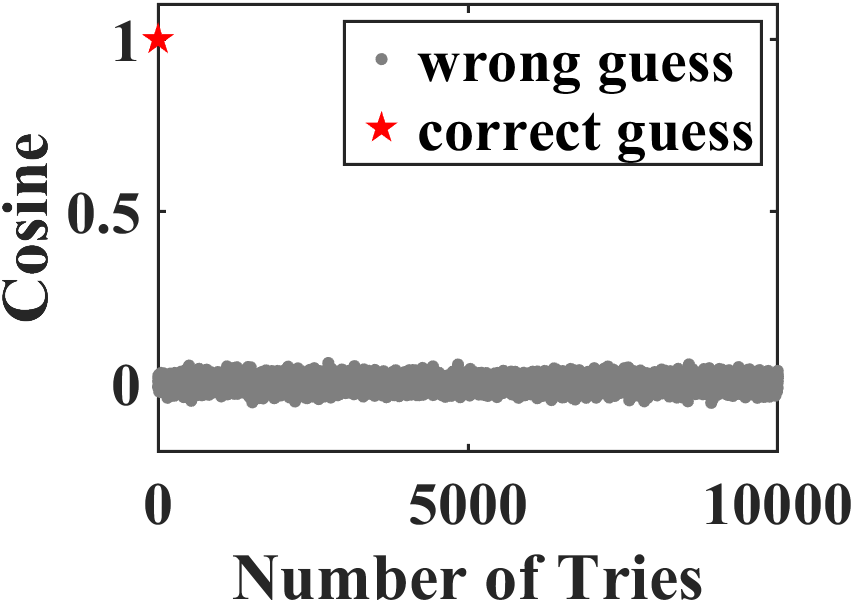}}
\subfigure[{\scriptsize Attack on $index(B_{1,2})$}]{
        \centering
        \label{fig:comb2_nonbin}
		\includegraphics[width=0.44\linewidth]{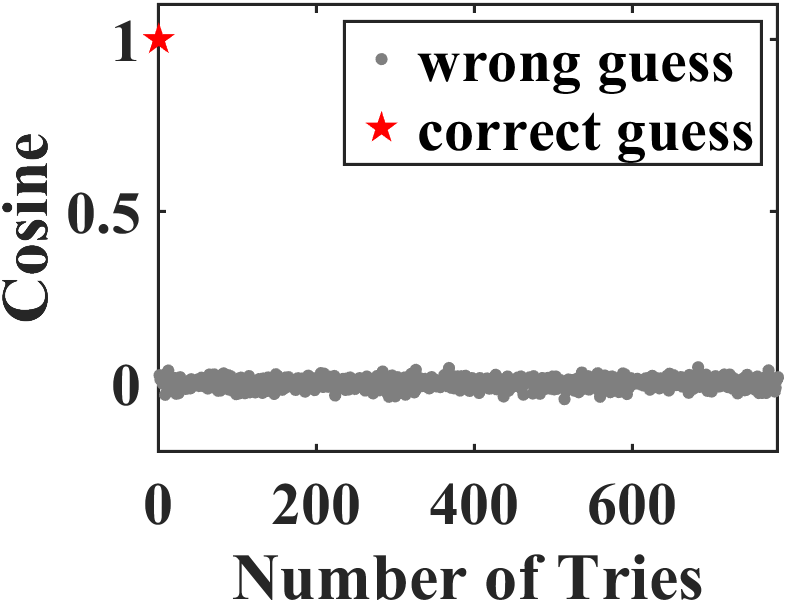}}
\caption{ \ouralg\ security validation on non-binary HDC.}
\label{fig:validation_nonb}
\end{minipage}%
\hspace{0.02\textwidth}
\begin{minipage}[h]{.3\textwidth}
\centering
\captionof{table}{The reasoning time for HDC models and the reconstructed model accuracy on different benchmarks}
\resizebox  {1.1\linewidth}{!}{
\begin{tabular}{ll|ccccc}
\toprule
\multicolumn{2}{c}{\textbf{Benchmark}} & \textbf{MNIST} & \textbf{UCIHAR} & \textbf{FACE} & \textbf{ISOLET} & \textbf{PAMAP} \\ \hline
\multicolumn{1}{l|}{\multirow{3}{*}{\begin{tabular}[c]{@{}l@{}}Non-\\ Binary \\ HDC \\ Model\end{tabular}}} & \begin{tabular}[c]{@{}l@{}}Original \\ Accuracy\end{tabular} & 0.8176 & 0.8385 & 0.9390 & 0.8839 & 0.8426 \\ \cline{2-2}
\multicolumn{1}{l|}{} & \begin{tabular}[c]{@{}l@{}}Recovered \\ Accuracy\end{tabular} & 0.8171 & 0.8381 & 0.9390 & 0.8845 & 0.8426 \\ \cline{2-2}
\multicolumn{1}{l|}{} & \begin{tabular}[c]{@{}l@{}}Reasoning\\ Time (s)\end{tabular} & 4057.59 & 1404.33 & 7388.32 & 1649.81 & 0.85 \\ \hline
\multicolumn{1}{l|}{\multirow{3}{*}{\begin{tabular}[c]{@{}l@{}}\\Binary \\ HDC\\ Model\end{tabular}}} & \begin{tabular}[c]{@{}l@{}}Original \\ Accuracy\end{tabular} & 0.7980 & 0.8164 & 0.9350 & 0.8685 & 0.8156 \\ \cline{2-2}
\multicolumn{1}{l|}{} & \begin{tabular}[c]{@{}l@{}}Recovered \\ Accuracy\end{tabular} & 0.7946 & 0.8181 & 0.9350 & 0.8724 & 0.8156 \\ \cline{2-2}
\multicolumn{1}{l|}{} & \begin{tabular}[c]{@{}l@{}}Reasoning\\ Time (s)\end{tabular} & 4284.27 & 1674.99 & 9100.14 & 2750.30 & 5.89 \\ \bottomrule
\end{tabular}}
\label{tab:reasoning_performance}
\end{minipage}
\end{figure*}

To explicitly show the robustness of \ouralg, we again attack the first pixel of MNIST binary HDC model, where $N=784$ and $D=10,000$. We set $P=N=784$ and $L=2$ for the validation\footnote{With $P=N$, these base hypervectors can directly serve as the feature hypervectors for the normal unprotected HDC model.}, so there are 4 parameters $\{k_{1,1}, index(B_{1,1}), k_{1,2}, index(B_{1,2})\}$ defining the encoding module. We assume in the worst case, that an adversary had successfully learned three parameters, and there is only one parameter left to be attacked. The Hamming distances of the guessed results are shown in Fig. \ref{fig:validation_binary}. For clear illustration, we plot the correct guess first and try all other wrong selections as followed. As the validation result shows, even if the attacker already correctly guessed most parameters, the mapping information will be useless as long as there is still one parameter incorrectly reasoned. The attacker has to apply $4.81\times 10^{16}$ tries to get the correct mapping information for the encoding module of the MNIST dataset. 
The validation on the non-binary HDC is shown in Fig. \ref{fig:validation_nonb}, where $cosine$ value is calculated to indicate the similarity between the guess and the golden reference. 
Similar to the \ouralg\ on binary HDC model, even if most parameters are correctly guessed, the derived mapping information will be useless as long as one of them is wrong. Hence, The number of needed tries is still $4.81\times 10^{16}$, which makes the reasoning on \ouralg\ significantly expensive.

\section{Experimental Evaluation} \label{sec:Evaluation}
In this section, we comprehensively evaluate the discovered vulnerability and the proposed locking strategy on several popular benchmarks, ranging from small to large dataset: MNIST (handwritten classification) \cite{726791}, UCIHAR (human activity) \cite{anguita2013public}, FACE (face recognition), ISOLET (voice recognition) \cite{Dua:2019}, PAMAP (physical activity) \cite{reiss2012introducing}. These datasets are commonly used for HDC performance analysis in previous works \cite{imani2019quanthd, imani2019searchd}. For the FACE dataset, we collect 623 face images in the open-source CMU Face Images dataset \cite{Dua:2019}, while 623 non-face images are randomly selected from the CIFAR-100 dataset \cite{krizhevsky2009learning} as a small-scale benchmark. All the experiments are evaluated with Python on an 3.60GHz Intel i7 processor with 16GB memory.

\subsection{Attacks on Different Benchmarks}
The experimental results shown in Tab. \ref{tab:reasoning_performance} compare the original (correct) HDC model and the reconstructed (estimated) HDC model by the reasoned hypervectors. For all the benchmarks, the reconstructed model can still achieve the same good accuracy as the original one, which means the mapping information of feature and value hypervectors are unfortunately leaked. Moreover, all the reasoning attacks can be completed in 3 hours, demonstrating the practical feasibility and efficiency of the reasoning attack for the normal HDC models.

\subsection{\ouralg\ Performance Evaluation}
\begin{figure}[!t]
\centering
\subfigure[Guesses v.s. $D$ and $P$]{
        \centering
        \label{fig:try_vs_DP}
		\includegraphics[width=0.5\linewidth]{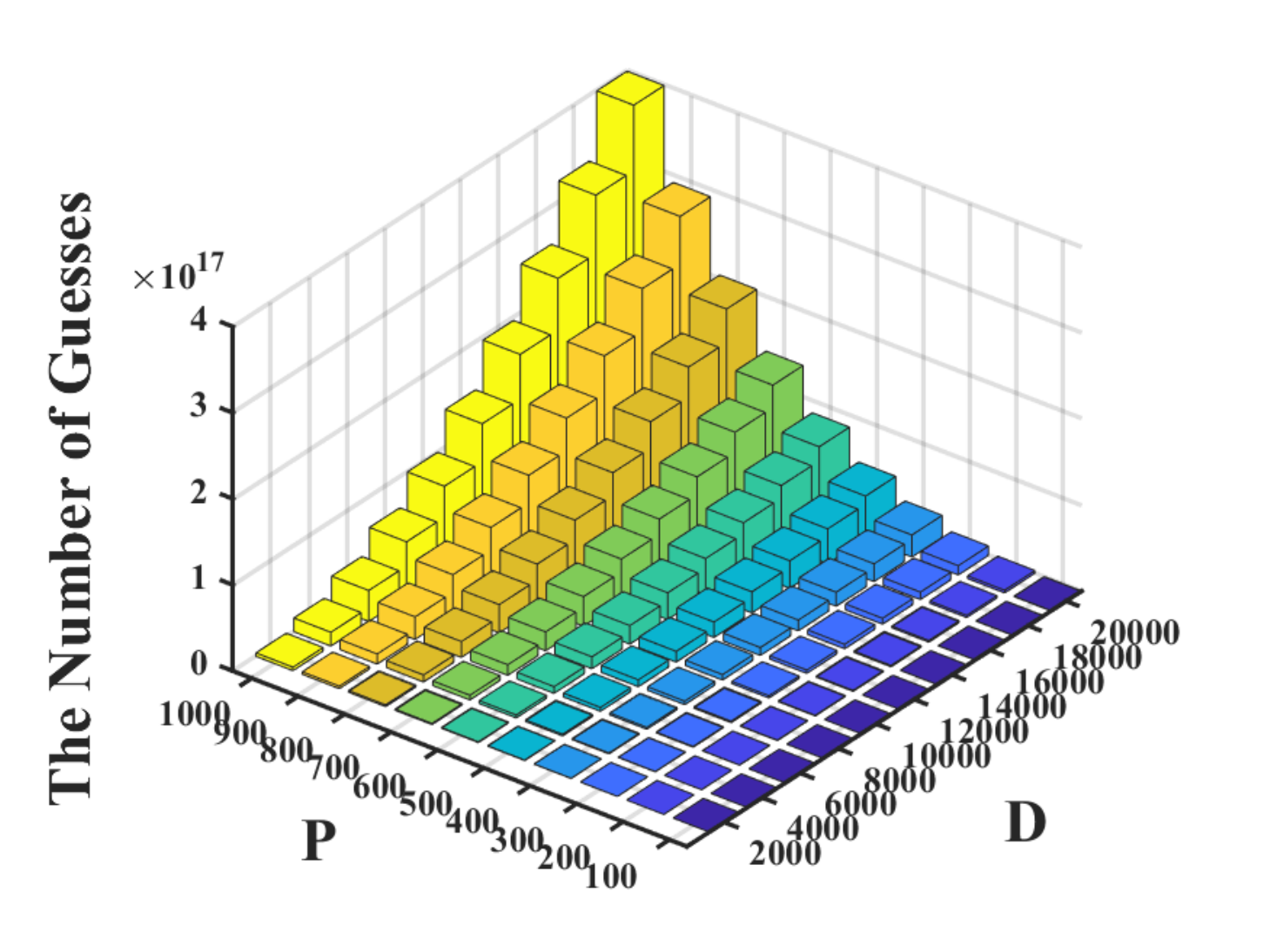}}
\subfigure[Guesses v.s. $L$ with various features]{
        \centering
        \label{fig:try_vs_L}
		\includegraphics[width=0.46\linewidth]{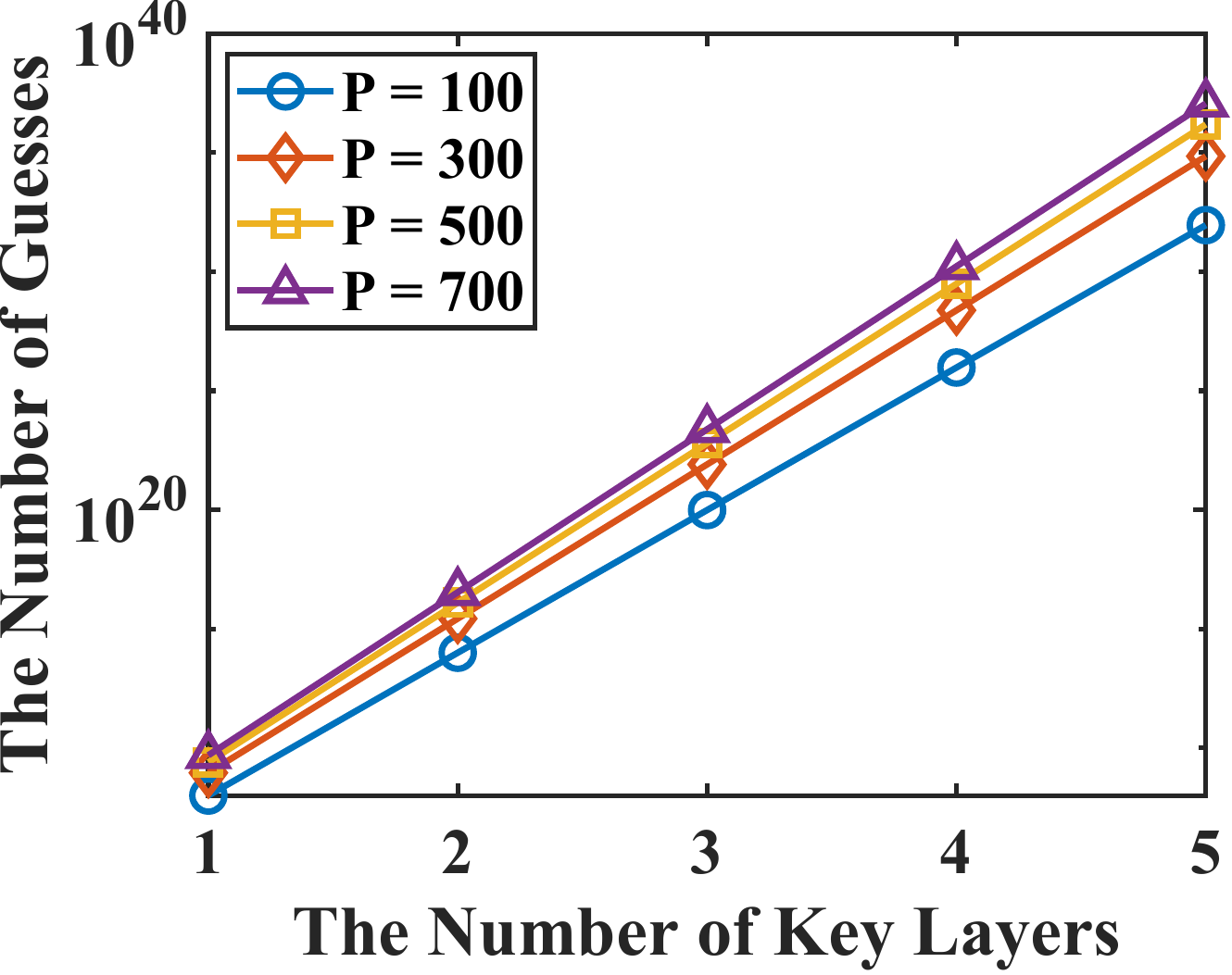}}

\caption{(a) The number of guesses versus the dimension $D$ and the number of base hypervectors $P$, assuming $L=2$. (b) The number of guesses versus the number of key layers $L$ with different number of base hypervectors $P$, assuming $D = 10,000$. Here the y-axis is in log-scale.}
\label{fig:try_vs_param}
\end{figure}

Compared to the normal HDC models, the permutation and combination schemes in the encoding phase by \ouralg\ can significantly increase the attacking complexity, from $O(N^2)$ to $O(N\cdot(DP)^{L})$. We show the number of needed reasoning guesses in theory in Fig. \ref{fig:try_vs_param}, which aligns with the time consumption if each guess costs approximately equal time. The number of guesses increases monomially with the power of $L$ ($L=2$ in our case), along with the $D$ and $P$ increments. Further, the number of guesses increases exponentially along with the key layers $L$, as shown in Fig. \ref{fig:try_vs_L}. Also, we demonstrate that $P$ and $L$ are mutually enhanced, i.e., increment on $P$ can introduce more complexity on the locking framework when $L$ is larger.

\begin{figure}[!t]
\centering
\subfigure[Non-binary record-based encoding]{
        \centering
        \label{fig:nb_acc_vs_L}
		\includegraphics[width=0.48\linewidth]{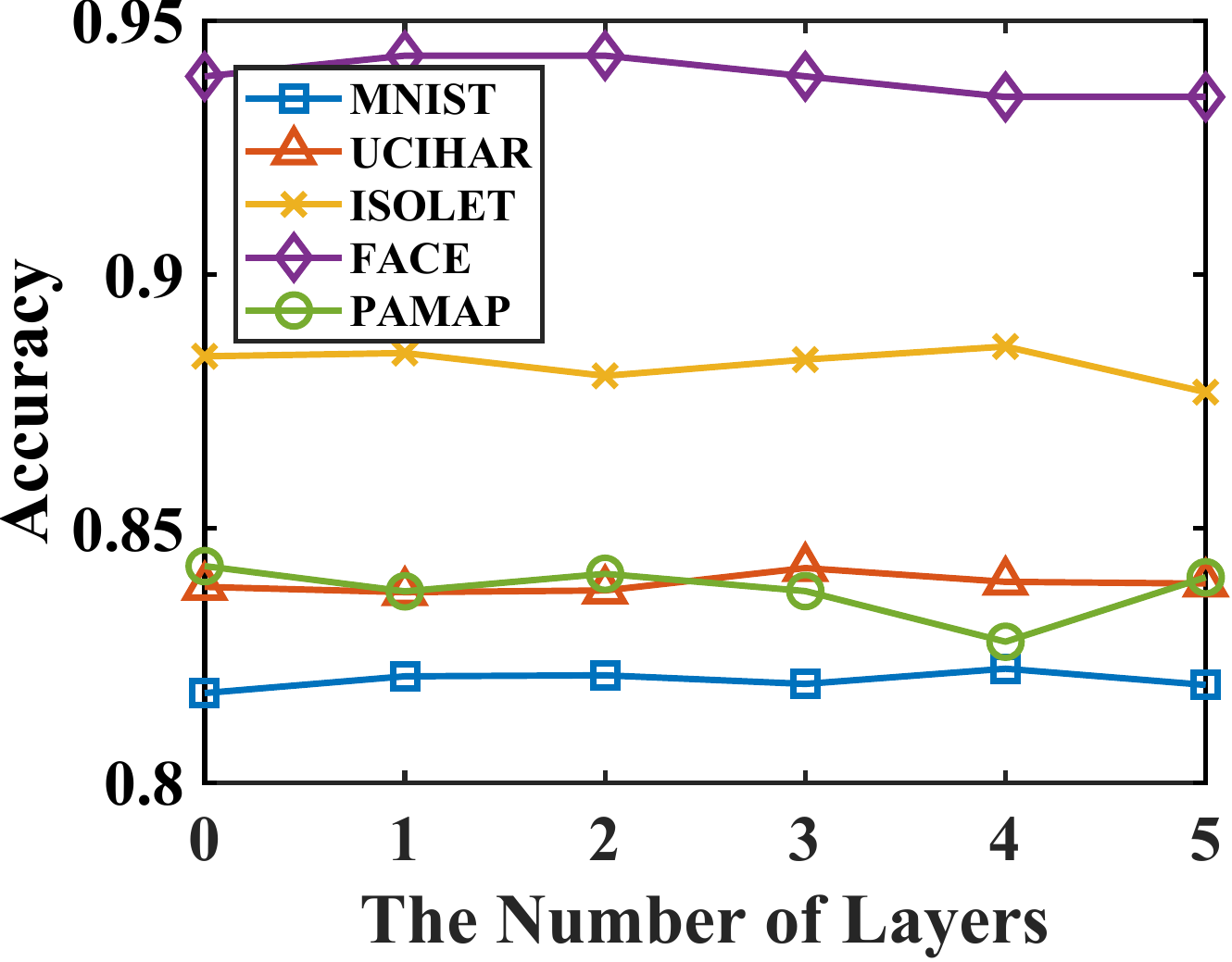}}
\subfigure[Binary record-based encoding]{
        \centering
        \label{fig:b_acc_vs_L}
		\includegraphics[width=0.48\linewidth]{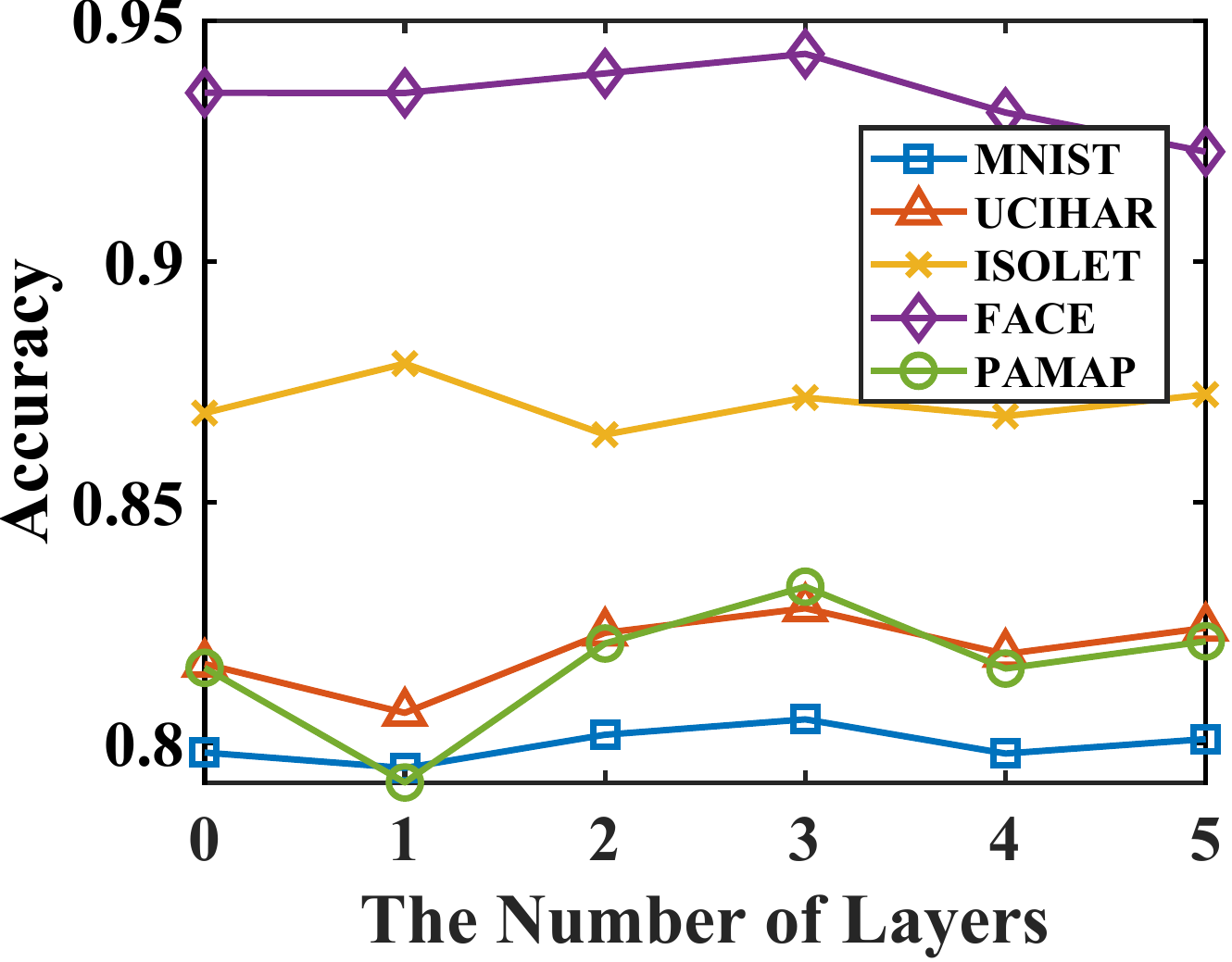}}

\caption{The accuracy changing for benchmarks on (a) non-binary and (b) binary record-based encodings. Here $L=0$ means the baseline HDC model without protection.}
\label{fig:acc_vs_L}
\end{figure}

Note that there exists trade-off while choosing the number of layers $L$, as more layers will lead to longer encoding time, while less layers might not ensure the security against powerful computing resources. To analyze such trade-off on practical hardware setup, we deploy the \ouralg\ framework on a Xilinx Zynq UltraScale+ FPGA \cite{zynq_ultrascale}. The HDC computing is segmented, pipelined and paralleled as tree structure, as discussed in \cite{imani2019quanthd}. 

Fig. \ref{fig:acc_vs_L} shows the accuracy comparison between the baseline (non-)binary HDC model and the \ouralg\ framework with different number of layers. The result demonstrates that there is no observable negative impact on the accuracy while applying \ouralg. This is because the encoding module of \ouralg\ does not change the orthogonality of feature hypervectors or the correspondence between the encoding input and output.

\begin{figure}[!t]
  \centering
  \includegraphics[width=0.45\linewidth]{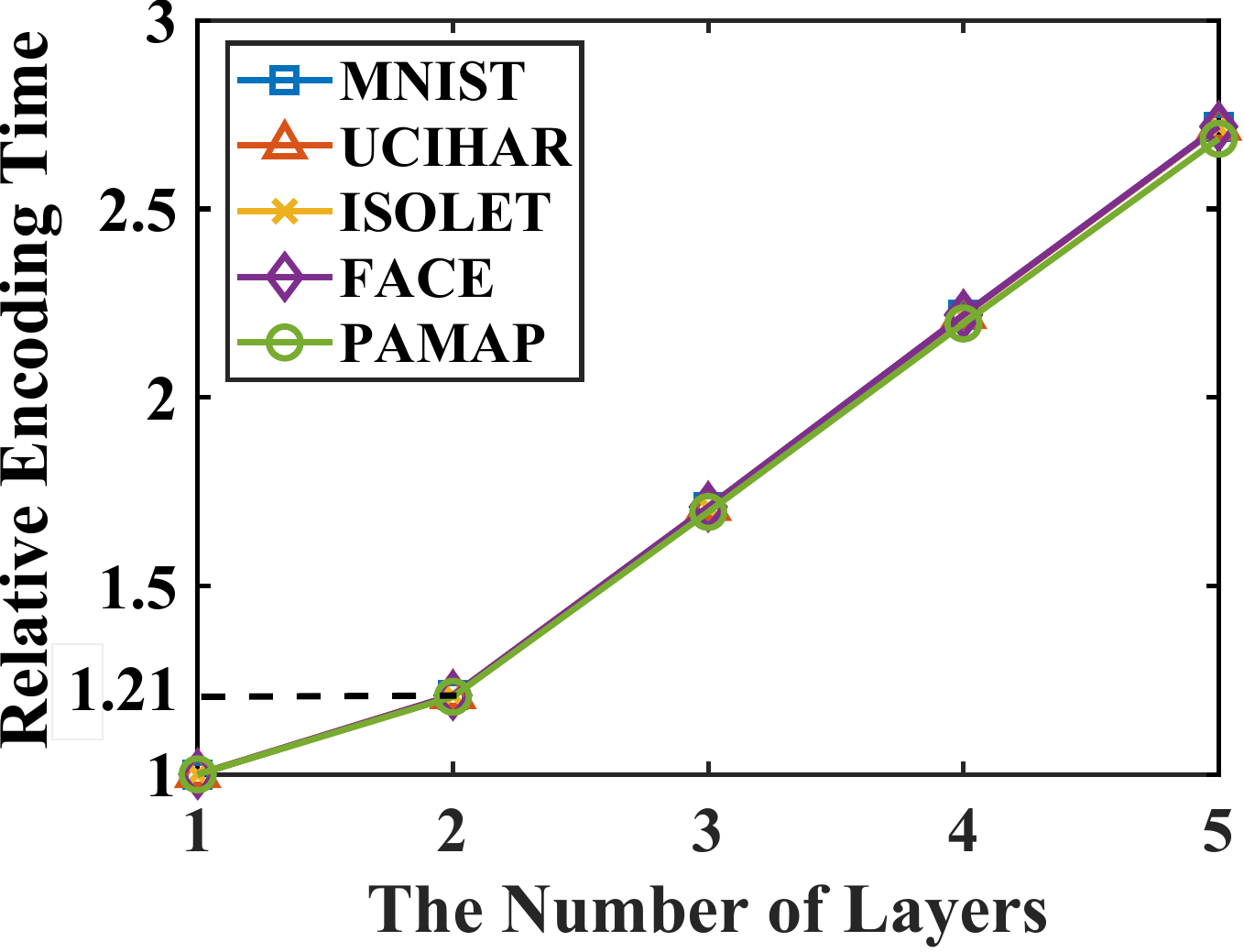}
  \caption{The encoding time changing of \ouralg\ framework relative to baseline HDC model. To give objective comparison, clock cycles are utilized as the encoding time, so the relative encoding time is the ratio of two clock-cycle measurements.}
  \label{fig:performance}
\end{figure}

Then we evaluate the time overhead caused by the introduced combination and permutation in \ouralg. Since only the encoding procedure is different between \ouralg\ and the baseline HDC model, we compare the encoding time overhead to provide objective and representative comparison. Also, we combine the non-binary and binary HDC model comparison, since they have identical encoding except for the binarization. The experimental results in Fig. \ref{fig:performance} show that the encoding time increases relative to the baseline HDC models. Besides, an important observation is that the increasing trend of all datasets almost coincide, which shows the encoding time growth is  independent of the dataset scale, as long as the hardware resource is sufficient. On the other hand, for $L=1$ (single-layer \ouralg), the relative encoding time is 1. In this scenario, a feature hypervector ($FeaHV$) is directly generated from a permuted base hypervector ($\rho(B)$), while the permutation is equivalent to shifted memory access, which will not require extra calculation compared to directly accessing the memory. From $L=2$, the encoding time increases linearly. 

Taking the MNIST dataset as an example, the one-layer key can provide $6.15\times 10^9$ attacking complexity while not consuming extra latency. Moreover, the two-layer key can provide $4.81\times 10^{16}$ attacking complexity, compared to $6.15\times 10^5$ in normal HDC models (i.e., $7.82\times 10^{10}$ times improvement), while only requiring 21\% encoding time overhead, as shown in Fig. \ref{fig:performance}. As a result, the reasoning cost makes it infeasible for an attacker with current computing resources to steal the IP of an HDC model. 

\section{Conclusion} \label{sec:Conclusion}
In this paper, we raise an urgent vulnerability of the emerging HDC models, which could be utilized by the adversary to steal the IP of the critical encoding module. To mitigate this vulnerability, we present a defense framework, \ouralg, to protect the encoding module. With \ouralg, the combination and permutation of multiple base hypervectors are used to generate hypervectors representing features. Experimental evaluations demonstrate that, the security of the encoding module protected by \ouralg\ is exponentially increased without incurring inference accuracy loss, while the time overhead is only linearly increased. With a two-layer key protection, \ouralg\ can increase the security by 10 order of magnitudes, while only consumes 21\% extra latency.

\input{ref.bbl}

\end{document}